\documentclass[pre,twocolumn,showpacs]{revtex4}
\usepackage{graphicx}

\newcommand{\equ}{Eq.}

\newcommand{\rem}[1]{}

\newcommand{\real}[1]{\text{Re}(#1)}

\begin{document}

\title{Simulating noisy quantum protocols 
with quantum trajectories} 
\author{Gabriel G. Carlo, Giuliano Benenti, Giulio Casati, and Carlos 
Mej\'\i a-Monasterio}
\email{gabriel.carlo@uninsubria.it}
\homepage{http://www.unico.it/~dysco}
\affiliation{Center for Nonlinear and Complex Systems, Universit\`a degli 
Studi dell'Insubria and Istituto Nazionale per la Fisica della Materia, 
Unit\`a di Como, Via Valleggio 11, 22100 Como, Italy}
\date{\today}

\begin{abstract} 
The theory of quantum trajectories is applied to simulate the effects
of quantum noise sources induced by the environment 
on quantum information protocols. 
We study two models that generalize single qubit noise channels like 
amplitude damping and phase flip to the many-qubit situation. 
We calculate the fidelity of quantum information transmission 
through a chaotic channel using the teleportation scheme with 
different environments. 
In this example, we analyze the role played by the kind of collective 
noise suffered by the quantum processor during its operation.
We also investigate the stability of a quantum algorithm simulating 
the quantum dynamics of a paradigmatic model of chaos, the baker's map.
Our results demonstrate that, using the quantum trajectories approach, 
we are able to simulate quantum protocols in the presence of noise and 
with large system sizes of more than 20 qubits.
\end{abstract}
\maketitle
										
\section{Introduction}
\label{sec:intro}

It would be highly desirable to implement quantum protocols 
using processors perfectly isolated from the environment, since 
this is one of the main sources of error in quantum computation 
(there are also system specific imperfections, but here we will 
only address the environmental problem).  
Unfortunately, this is not possible. Quantum hardware will 
naturally become entangled with the environment during its operation. 
Thus, if any hope of profiting from the benefits of quantum computation 
is to be kept, understanding and controlling quantum noise effects 
is essential. On the other hand, the study of open systems is of 
interest in several fields, from both theoretical and experimental 
points of view \cite{Gra1,Gra2,Gra3,Knight}.  

Factoring large integers in polynomial time has been the milestone 
discovery that set out quantum computation as a major research topic
\cite{shor}. 
Nevertheless, in the short term, few qubits quantum computers, 
this kind of calculations will be necessarily out of reach, 
since they involve large systems. Then, it is reasonable to focus on 
understanding the behavior of accessible first realizations. 
It is interesting to remark that, with a few tens 
of qubits, quantum simulations of systems studied 
in quantum chaos (like the quantum baker's map \cite{Sch}, 
the quantum kicked rotator \cite{Bertrand}, and the quantum sawtooth 
map \cite{Ben1}) would outperform any calculation that can be done 
with present day supercomputers. 
A first step in the simulation of quantum chaos models has been
the implementation of the quantum baker's map on a three-qubit
NMR-based quantum processor \cite{cory}.

With this situation in mind, it is natural to ask to what extent we 
can know and control the operability and the stability of a 
quantum computer of this size. Theoretical studies rely 
on evaluating the reduced density matrix of the system 
(obtained after tracing out the environment), often in 
terms of various different approximations.
It would therefore be desirable to give an answer for generic
quantum protocols and noise models, using exact calculations.  
In this work, we propose the numerical simulation of superoperators 
as a way to do it. By means of quantum trajectories techniques 
we can reach results for system sizes for which the implementation 
of chaotic maps becomes relevant for theoretical studies in the 
field of quantum chaos, having the chance to include any kind of 
environmental effects. 

Instead of solving the density matrix directly, quantum trajectories 
stochastically evolve the state vector of the system, 
and after averaging over many runs the same results for 
the outcomes of any observable are obtained. 
The use of quantum trajectories in the field of quantum information
has been pioneered by \cite{Schack,BarencoBrun}.
In Ref.~\cite{Car}, we applied this formalism to study the 
effect of a dissipative environment 
on the quantum  teleportation protocol \cite{BennettBrassard} 
through a large chain of qubits.
This situation models the transmission of quantum information 
through a chaotic quantum channel.
Here we review this model and present new results for different 
noise channels. 

As mentioned before, one of the main near future applications 
of quantum computation is the simulation of dynamical systems 
that are of great interest in quantum chaos.
In this work we focus on the quantum baker's map, one of the 
most important examples of this kind of systems. 
This map is fully chaotic and its quantized version consists of 
conveniently selected quantum Fourier transforms. 
We model the environment through a phase flip channel 
and study the fidelity of quantum computing of the quantum
baker's map in this noisy environment. We note that 
the fidelity has already been computed experimentally with 
a three-qubit NMR quantum processor \cite{cory}.
Hence, for the design and construction of quantum 
hardware with a larger number of qubits, simulations 
like those performed in this paper will become essential. 

This paper is organized as follows. In Sec.~\ref{sec:QT}, we present 
a brief explanation of the theory of quantum trajectories and connect 
it with the quantum operations approach to the density matrix 
evolution. We also refer to the master equation formulation. 
Afterward, in Sec.~\ref{sec:noise}, we make a short description 
of the noise channels that we use in the calculations, paying special 
attention to the amplitude damping models. In Sec.~\ref{sec:tele}, 
we review the quantum teleportation protocol presented in \cite{Car}, 
focusing on the different dissipative processes. 
In Sec.~\ref{sec:Baker}, we study the behavior of the fidelity
of the quantum algorithm for the baker's map in the presence 
of a phase flip noise. 
Finally, in Sec.~\ref{sec:conclusion}, we present our 
conclusions and outlook.

\section{Master equation, superoperators, and quantum trajectories}
\label{sec:QT}

There is a close relationship among the master equation in Lindblad 
form, the superoperator formalism and the quantum trajectories theory. 
It is useful to review this relationship and to motivate the use of the 
quantum trajectory approach when studying open systems 
like quantum processors. 
We start with the special case of the master equation formulation 
of the problem and relate it with simulations using quantum trajectories 
techniques. Then, in Sec.~\ref{sec:QT:qops} we generalize this 
establishing the connection with the broader quantum operators formalism. 

\subsection{Master equation and quantum trajectories}
\label{sec:QT:meqt}

Real systems interact with the environment, 
and, as mentioned before, these interactions are usually referred 
to as {\em quantum noise} \cite{Chuangbook,Preskillbook}. 
Several models account for different sorts of system-environment
interactions, the particular choice depending on the nature of 
the system and environment under consideration. 
Such open quantum systems in general cannot be described by a pure state, 
but rather by a mixed state. 
Their evolution takes density matrices to density matrices. 
This allows the evolution from system's pure states to mixed ones, 
and also, in some noise models, from mixed to pure states. 
In order to obtain the differential equation corresponding to this 
process, one assumes Markovian behavior, giving the evolution of the 
density operator with reference only to its state at present. 
This Markovian assumption neglects memory effects: it implies that 
the time needed for the environment to loose the information it 
received from the system is short enough in comparison with the time 
scale of the dynamics we perceive. Then, we are entitled to regard the 
information flow in only one direction, neglecting any kind of 
feedback \cite{Preskillbook}.

The Lindblad form of the master equation of this system-environment 
model in the Born-Markov approximation can be formally written as 
\cite{Lindblad}:
\begin{equation}
\dot \rho = {\cal L} [\rho], 
\end{equation}
with formal solution
\begin{equation}
\rho(t) = \exp({\cal L}t) [\rho(0)], 
\end{equation}
where ${\cal L}$ stands for the ``Lindblandian'' operator. 
In order to obtain explicit expressions we can trace out the 
environment, which gives
\begin{equation}
\dot \rho = -\frac{i}{\hbar} [H_s,\rho] - \frac{1}{2} \sum_{\mu} 
\{L_{\mu}^{\dag} L_{\mu},\rho\}+\sum_{\mu} L_{\mu} \rho 
L_{\mu}^{\dag},
\label{eq:lin2}
\end{equation}
where $L_{\mu}$ are the Lindblad operators 
($\mu \; \epsilon \; [1,\ldots,{\cal M}]$, the number ${\cal M}$
depending on the noise model), 
$H_s$ is the system's Hamiltonian
and \{\,,\,\} denotes the anticommutator. 
The first two terms of this equation can be regarded as the evolution 
performed by an effective non-Hermitian Hamiltonian, $H_{\rm eff}=
H_s+iK$, with $K=-\hbar/2 \sum_{\mu}L_{\mu}^{\dag}L_{\mu}$. 
In fact, we see that 
\begin{equation}
 -\frac{i}{\hbar} [H_s,\rho] - \frac{1}{2} \sum_{\mu} 
\{L_{\mu}^{\dag} L_{\mu},\rho\} = -\frac{i}{\hbar} 
[H_{\rm eff} \rho - \rho H_{\rm eff}^{\dagger}],
\end{equation}
which reduces to the usual evolution equation for the 
density matrix in the 
case of $H_{\rm eff}$ being Hermitian.
The last term is responsible for the 
so-called {\em quantum jumps}. 
In this context the Lindblad 
operators $L_{\mu}$ are also named {\em quantum jump operators}. 
If the initial density matrix is in a pure state 
$\rho(t_0)=|\phi(t_0)\rangle \langle\phi(t_0)|$, 
after a time $dt$ evolves to the following statistical mixture: 
\begin{equation}
\rho(t_0+dt)=(1-\sum_{\mu}dp_{\mu}) 
\: |\phi_0\rangle \langle\phi_0| \: + 
\sum_{\mu} dp_{\mu}  |\phi_{\mu}\rangle \langle\phi_{\mu}|, 
\label{eq:stat}
\end{equation}
with the probabilities $dp_{\mu}$ defined by 
\begin{equation}
dp_{\mu}\!=\!\langle \phi(t_0)| L_{\mu}^{\dag} 
L_{\mu} |\phi(t_0)\rangle dt,
\end{equation} 
and the new states by 
\begin{equation}
|\phi_0\rangle = 
\frac{(\openone-i H_{\rm eff} dt/\hbar) |\phi(t_0)\rangle}
{\sqrt{1-\sum_{\mu} dp_{\mu}}} 
\label{eq:stat2}
\end{equation}
and
\begin{equation}
 |\phi_{\mu}\rangle = \frac{L_{\mu} 
|\phi(t_0)\rangle}{||L_{\mu} |\phi(t_0)\rangle||}. 
\end{equation}

Then, the {\em quantum jump picture} turns out to be clear; 
with probability 
$dp_{\mu}$ a jump occurs and the system is prepared in the state 
$|\phi_{\mu}\rangle$. With probability $1-\sum_{\mu} dp_{\mu}$ there 
are no jumps and the system evolves according to the effective 
Hamiltonian $H_{\rm eff}$ (normalization is included also in this case 
because the evolution is given by a non-unitary operator).

The numerical method we are going to use in order to 
simulate the master equation is usually known as the Monte Carlo 
Wave function approach \cite{DalibardCastin}. 
We start from a pure state $|\phi(t_0)\rangle$ 
and at intervals $dt$, smaller than 
the timescales relevant for the evolution of the density matrix, 
we perform the following evaluation. 
We choose a random number $\epsilon$ 
from a uniform distribution in the unit interval $[0,1]$. 
If $\epsilon < dp$, where $dp=\sum_{\mu} dp_{\mu}$, 
the system jumps to one of the states $|\phi_{\mu}\rangle$ 
(to $|\phi_1\rangle$ if $0 \le \epsilon \le dp_1$, to 
$|\phi_2\rangle$ if $dp_1 < \epsilon \le dp_1+dp_2$, and so on). 
On the other hand, if $\epsilon > dp$, the evolution with 
the non-Hermitian Hamiltonian $H_{\rm eff}$ takes place, ending 
up in the state $|\phi_0\rangle$. In 
both circumstances we 
renormalize the state. We repeat this process as many times as 
$n_{\rm steps}=\Delta t/dt$ where $\Delta t$ is the whole 
elapsed time during the evolution. 
Note that we must take $dt$ much smaller than the time scales relevant
for the evolution of the open quantum system under investigation.
In our simulations, $n_{\rm steps}$ will be 
proportional to the number of quantum gates involved in 
the corresponding protocol. Each realization 
provides a different {\em quantum trajectory} and a particular set 
of them (given a choice of the Lindblad operators) is an 
``unraveling'' of the master equation. 
It is easy to see that if 
we average over different runs we recover the probabilities 
obtained with the density operator. In fact, given an operator 
$A$, we can write the mean value $\langle A \rangle_t={\rm Tr}
[A \rho(t)]$ as the average over ${\cal N}$ trajectories: 
\begin{equation}
\langle A \rangle_t = 
\lim_{{\cal N}\to \infty}
\frac{1}{{\cal N}} \sum_{i=1}^{{\cal N}} \langle \phi_i(t)| A 
| \phi_i(t) \rangle.
\end{equation}

The advantage of using the quantum trajectories method is clear 
since we need to store a vector of length $N$ ($N=2^n$ is 
the dimension of the Hilbert space, $n$ being the number
of qubits) rather than a $N \times N$ density matrix. 
Moreover, there is also an advantage in computation 
time with respect to density matrix direct calculations. 
We find that a reasonable 
amount of trajectories (we have used $100\le {\cal N} \le 400$ in 
all calculations, unless otherwise mentioned) is needed in 
order to obtain a satisfactory statistical convergence. 

This picture can be formalized by 
means of the stochastic Schr\"odinger equation \cite{Brun}
\begin{eqnarray}
|d \phi \rangle & = & -i H_s |\phi\rangle dt - \frac{1}{2} \sum_{\mu} 
(L_{\mu}^{\dag} L_{\mu} - \langle L_{\mu}^{\dag} L_{\mu} 
\rangle_{\phi}) |\phi\rangle \: dt \nonumber  \\ 
 & & + \sum_{\mu} 
\left( \frac{L_{\mu}}{\sqrt{\langle L_{\mu}^{\dag} L_{\mu} 
\rangle_{\phi}}}
-\openone \right) |\phi \rangle \: dN_{\mu}.
\label{stocsch}
\end{eqnarray}  
This is a stochastic nonlinear differential equation, where the 
stochasticity is due to the measurement results: 
we think that the environment is actually measured (as it is 
the case in indirect measurement models) or simpler, that 
the contact of the system with the environment produces an 
effect similar to a continuous measurement \cite{Zurek}. 
The nonlinearity is due to the renormalization of the state 
vector. 
The stochastic differential variables $dN_{\mu}$ are 
statistically independent and represent measurement outcomes. 
Their ensemble mean is given by 
$M[dN_{\mu}]=\langle L_{\mu}^{\dag} L_{\mu} \rangle_{\phi} dt$. 
The probability that the variable $dN_\mu$ 
is equal to $1$ during a given time step $dt$
is $\langle \phi L_{\mu}^{\dagger} L_{\mu} \phi \rangle \: dt$.
Therefore, most of the time the variables 
$dN_{\mu}$ are $0$ and as a consequence the system evolves 
continuously by means of the non Hermitian effective Hamiltonian. 
However, when a variable $dN_{\mu}$ is equal to $1$,
the corresponding term in equation (\ref{stocsch}) 
is the most significant. In these cases the quantum jump occurs. 
Note that there are also other possibilities in order to unravel 
the master equation such as the quantum state diffusion 
\cite{GisinPercival,Schack}, for example.

\subsection{Quantum operations and quantum trajectories}
\label{sec:QT:qops}

There is a close connection between the master equation in 
Lindblad form and the quantum operations theory 
\cite{Chuangbook,Preskillbook} and 
we use this latter formalism as a comparison tool for the quantum 
trajectories calculations. 
Furthermore, it will become clear that the evolution 
of the density matrix of the system given by this method can 
be put on the same 
footing as the stochastic evolution model. In fact, the latter 
constitutes a Monte Carlo simulation of the former. 

We write the solution to 
\equ~(\ref{eq:lin2}) over an infinitesimal time $dt$
as a completely positive map:
\begin{equation}
\rho(t+dt)=\$[\rho(t)]=\sum_{\mu=0}^{{\cal M}} M_{\mu}(dt) \rho(t) 
M_{\mu}^{\dag}(dt),
\label{eq:infsolution}
\end{equation}
where, for $\mu=0$, we have $M_0=\openone - i H_{\rm eff} dt/\hbar$ and, for 
$\mu > 0$, $M_{\mu}=L_{\mu} \sqrt{dt}$, satisfying $\sum_{\mu=0}^{\cal M} 
M_{\mu}^{\dag} M_{\mu} = \openone$ to first order in $dt$. 
Equation (\ref{eq:infsolution}) is called the Kraus representation
(or the operator sum representation) of the superoperator $\$$
and the operators $M_\mu$ are known as operators elements
for the quantum operation $\$$.
It can be shown that, if the global evolution (system 
plus environment) is unitary, the Kraus operators 
satisfy the completeness relation $\sum_{\mu=0}^{\cal M} 
M_{\mu} M_{\mu}^{\dag} = \openone$ \cite{Chuangbook,Preskillbook}. 
Note that the superoperator $\$$ maps density matrices to density 
matrices, that is $\rho(t+dt)$ is Hermitian, has unit trace and 
is nonnegative if $\rho(t)$ satisfies these properties. 

The action of the quantum operation $\$$
can be interpreted as $\rho$ being randomly replaced 
by $M_{\mu} \rho M_{\mu}^{\dag} / {\rm Tr} (M_{\mu} \rho M_{\mu}^{\dag})$, 
with probability ${\rm Tr} (M_{\mu} \rho M_{\mu}^{\dag})$. Equivalently, 
the set $\{M_{\mu}\}_{\mu=1,...,{\cal M}}$ 
defines a Positive Operator Valued Measurement 
(POVM) with operators $E_{\mu}=M_{\mu}^{\dag} M_{\mu}$, 
that satisfy 
$\sum_{\mu=0}^{\cal M} E_{\mu}=\openone$. 
The outlined process is equivalent to performing a continuous 
measurement on the system (or an indirect measurement if the environment 
is actually measured) and shows the close
connection between the Kraus operators 
formalism and the quantum jumps picture. 
%

We would like to mention that the quantum operations formalism 
is more general than the master equation approach. 
The operator sum formulation in differential form can 
be obtained from the master equation, but a general quantum process 
described in terms of an operator sum representation needs to 
be Markovian in order to be tractable with a master equation. 
This opens the possibility of studying a wide range of 
non Markovian phenomena using quantum trajectories simulations. 


\section{Noise channels}
\label{sec:noise}

There are several ways to model the interaction of a system 
with the environment. The most common examples found in the literature 
are the amplitude damping channel, the phase flip channel,
and the depolarizing channel 
\cite{Chuangbook,Preskillbook}.
In this section, we give a brief description of two relevant noise
models (amplitude damping and phase damping) in the single qubit 
case, and then we generalize them to $n$-qubit systems.

Dissipation (energy loss) is one of the main features present 
in open quantum systems. Different phenomena like, for example, 
spontaneous atomic emission of a photon or spin systems approaching 
equilibrium can be modeled by this quantum operation, i.e., 
the amplitude damping. 
Considering the environment initially in the vacuum state 
$|0\rangle_e$, there is a probability $p$ that the excited state 
of the system decays and that the state of the environment 
changes from the vacuum to $|1\rangle_e$.  
\begin{eqnarray}
|0\rangle_s |0\rangle_e & \rightarrow & |0\rangle_s |0\rangle_e, 
\nonumber \\ 
|1\rangle_s |0\rangle_e & \rightarrow & \sqrt{1-p} \; |1\rangle_s, 
|0\rangle_e
+\sqrt{p} \; |0\rangle_s |1\rangle_e,
\label{damp1}
\end{eqnarray}
where $|0\rangle_s$ stands for the spin down (ground) state 
and $|1\rangle_s$ for the spin up (excited) state of the system,
and the indexes $s$ and $e$ denote the quantum states of the 
system and of the environment, respectively.
Tracing out the environment the corresponding Kraus operators are 
obtained: 
\begin{equation}
M_0= \left(
\begin{array}{cc}
1 & 0 \\
0 & \sqrt{1-p} \\
\end{array} \right), \quad
M_1 = \left(
\begin{array}{cc}
0 & \sqrt{p} \\
0 & 0 \\
\end{array} \right).
\label{Kraus1}
\end{equation}
The operator $M_1$ is responsible for the  
quantum jumps and $M_0$ for the continuous evolution. 
For the case of infinitesimal evolution operators  
(see Sec.~\ref{sec:QT}), a 
repeated application of this noise channel 
gives an exponential decay law
of the population of the state $|1\rangle$ \cite{Preskillbook}. 
Therefore, this evolution drives any intial (pure or mixed) 
state of the qubit to the pure state $|0\rangle$. 
Note that here and in the following the matrix representations 
of the single-qubit Kraus operators are written in the 
$\{|0\rangle,|1\rangle\}$ basis.

We also consider the phase flip channel (which is equivalent to 
the phase damping channel \cite{Chuangbook}) 
given by the following model
\begin{eqnarray}
|0\rangle_s |0\rangle_e & \rightarrow & \sqrt{1-p} \; |0\rangle_s 
|0\rangle_e + \sqrt{p} \; |0\rangle_s |1\rangle_e, \nonumber \\ 
|1\rangle_s |0\rangle_e & \rightarrow & \sqrt{1-p} \; |1\rangle_s 
|0\rangle_e - \sqrt{p} \; |1\rangle_s |1\rangle_e,
\end{eqnarray}
with Kraus operators
\begin{equation}
M_0=\sqrt{1-p} \; \left(
\begin{array}{cc}
1 & 0 \\
0 & 1 \\
\end{array} \right), \quad
M_1 = \sqrt{p} \; 
\left(
\begin{array}{cc}
1 & 0 \\
0 & -1 \\
\end{array} \right).
\label{Kraus2}
\end{equation}
This kind of noise can be thought as describing quantum 
information loss, in contrast with the 
previous model, which describes energy loss. 
This is a purely quantum mechanical process 
and has been extensively studied in the context of 
quantum to classical correspondence \cite{Zurek}.

In order to generalize the single qubit amplitude damping 
process to many qubits  we will follow two different 
points of view. 
In the first case we assume that a single damping probability 
describes the action of the environment, irrespective 
of the internal many-body state of the system. 
In the second approach, we assume that each qubit has 
its own interaction with the environment, independently of 
the other qubits. This makes the damping probability 
grow with the number of qubits that can perform the transition
$|1\rangle\to |0\rangle$.
Both models assume that only one qubit of the system can 
decay at a time. 

In the first case \cite{Car}, we have a probability $p$ 
for the system to perform one of the possible 
transitions, each of them being equally likely. 
This can be illustrated with a two-qubit example:
\begin{eqnarray}
|00\rangle_s |00 \rangle_e & \rightarrow & 
|00\rangle_s |00 \rangle_e, \nonumber \\
|01\rangle_s |00 \rangle_e & \rightarrow & 
\sqrt{1-p} \: |01\rangle_s |00 \rangle_e + 
\sqrt{p} \: |00\rangle_s |01\rangle_e, \nonumber \\
|10\rangle_s |00 \rangle_e & \rightarrow & 
\sqrt{1-p} \: |10\rangle_s |00 \rangle_e +
\sqrt{p} \: |00\rangle_s |10\rangle_e, \nonumber \\
|11\rangle_s |00 \rangle_e & \rightarrow & 
\sqrt{1-p} \: |11\rangle_s |00 \rangle_e +
\sqrt{p/2} \nonumber \\
 & & (|10\rangle_s |01\rangle_e + |01\rangle_s |10\rangle_e).
\end{eqnarray} 

A $n$-qubit state $|i_{n-1}\ldots i_0\rangle$ 
($i_l=0,1$, with $0\le j \le n-1$) decays in the interval $dt$ 
with a probability $p=\Gamma dt /\hbar$. 
After this infinitesimal time, the possible states of the system 
are those in which the damping $|1\rangle \to |0\rangle$ has 
occurred in one of the qubits, the damping probability being the same 
for all the qubits.
We provide a compact expression for 
the Kraus operators $M_{\mu}$ in our model, following the formulation 
given in Sec.~\ref{sec:QT:qops} for the infinitesimal evolution. 
The matrix elements for the $k$th operator 
in the computational basis 
$|i\rangle \equiv |i_{n-1}\ldots i_0\rangle$, with 
$i\equiv\sum_{l=0}^{n-1} i_l 2^l$, are given by   
\begin{equation}
[M_{\mu}^{(n)}]_{i,j} = \left\{
\begin{array}{ll}
 \sqrt{\frac{\Gamma dt}{\hbar \sum_{l=0}^{n-1} i_l}}, & 
\;{\rm for}\; j\geq 2^{(\mu-1)}, \\
 &  i=j-2^{(\mu-1)}, \\
 &  i_l=0,1, \\ 
 & \\
 0  & \;{\rm otherwise},
\end{array} \right.
\label{amp1}
\end{equation}
where the superscript ${}^{(n)}$ in $M_{\mu}^{(n)}$ 
underlines the fact that we are dealing with $n$-qubit
Kraus operators.
There are $n$ operators $M_{\mu}^{(n)}$ ($\mu=1,\ldots,n$), 
where the index $\mu$ singles out
which qubit undergoes the transition $|1\rangle \to |0\rangle$.
The $M_0^{(n)}$ operator is
\begin{equation}
[M_{0}^{(n)}]_{i,j} = \left\{
\begin{array}{ll}
1 & \; {\rm for}\; i=j=0, \\
 \sqrt{1-\frac{\Gamma dt}{\hbar}} & 
\;{\rm for}\; i=j \ne 0, \\ 
 & \\
 0  & \;{\rm otherwise}. 
\end{array} \right.
\end{equation}
Note that, if we replace in the above definition the square root by its 
first order approximation, we arrive at the same expression for 
the action of the effective Hamiltonian given 
in Sec.~\ref{sec:QT} (see the first term in the right hand side 
of \equ~(\ref{eq:stat}) and \equ~(\ref{eq:stat2})).

For example, starting from the four-qubit pure state
$\rho(t_0)=|1011\rangle\langle 1011|$, the action of the generalized 
amplitude damping channel leads, after a time $dt$, to 
the statistical mixture 
$$
\rho(t_0+dt)=\left(1-\frac{\Gamma dt}{\hbar}\right)
|1011\rangle\langle 1011|+
\frac{\Gamma dt}{3\hbar}(|0011\rangle\langle 0011|
$$
\begin{equation}
+|1001\rangle\langle 1001|+|1010\rangle\langle 1010|).
\label{ampex}
\end{equation}

\begin{figure}
\includegraphics[width=8.0cm,angle=0]{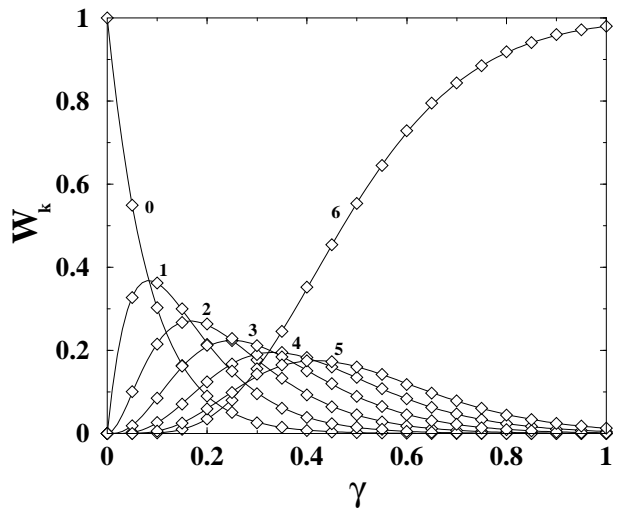}
\caption[]{\footnotesize Probabilities $W_k$ for each 
class of states in terms of the dimensionless damping rate 
$\gamma=\Gamma dt /\hbar$, 
in a system with $n=6$ qubits, obtained after evolving the initial
state $|1\ldots 1\rangle$ up to time $2n dt$, under the noise 
model (\ref{amp1}).
Solid lines correspond to the exact formulae (\ref{ADch0}), 
and the numbers close to each curve indicate the corresponding 
class. Diamonds 
stand for quantum trajectories simulations  
(error bars are not shown since they are smaller than the
size of the symbols).}
\label{fig:cascade}
\end{figure}

This situation can be described as {\em branching process} or 
a {\em cascade} in the population 
of different classes of states (see, e.g., Ref.~\cite{Flam}). 
A class in this model is naturally defined as the collection of 
all the states of the system having the same number of
qubits in the ``up'' state (i.e., of $|1\rangle$ states).  
The first class is the one initially populated with probability $W_0$.
The second class, with associated probability $W_1$, corresponds to the 
states that are obtained from the initial state after the damping of 
one qubit. The third class, with probability $W_2$, includes 
the states that can be reached by the damping of one qubit 
in any of the states of the second class, and so on until reaching the 
``ground state'' $|0\ldots 0\rangle$ of the system. 
This last state is also the only state that resides in  
the last class. This class is populated with probability 
$W_m$, the number $m$ being the maximum number of up spins in the 
initial state (if we start from a state 
$|i_{n-1}\ldots i_0\rangle$ of the computational basis, 
then $m=\sum_{l=0}^{n-1} i_l$). Note that $m\le n$. 
The set of probabilities $W_m$ 
can be obtained by means of the following differential equations:
\begin{eqnarray}
\frac{dW_0}{dt} & = & -\frac{\Gamma}{\hbar} \; W_0, \nonumber \\
 & \vdots & \nonumber \\
\frac{dW_k}{dt} & = & \frac{\Gamma}{\hbar} \; (W_{k-1} -  W_k), 
\nonumber \\
 & \vdots & \nonumber \\
\frac{dW_m}{dt} & = & \frac{\Gamma}{\hbar} \; W_{m-1},
\end{eqnarray}  
with solutions
\begin{eqnarray}
W_k & = & \frac{(\Gamma t /\hbar)^k}{k!} \; 
\exp{\left(-\frac{\Gamma t}{\hbar}\right)},
\nonumber \\
W_m & = & 1 - \sum_{k=0}^{m-1} W_k.
\label{ADch0}
\end{eqnarray}
In Fig.~\ref{fig:cascade} we show a comparison among the probabilities 
for each class obtained with quantum trajectories simulations and 
the theoretical formulae, in the case of $n=6$.
We have evolved the initial state $|2^n-1\rangle=|1\ldots 1\rangle$ 
up to time $2n \; dt$, for different values of the
dimensionless damping rate 
$\gamma=\Gamma dt/\hbar$. 
This evolution is purely   
dissipative, without considering any other kind of system's  
dynamics. As can be seen from Fig.~\ref{fig:cascade}, 
we have a very good agreement between quantum 
trajectories numerical simulations and the theoretical 
predictions of the model. 

The other generalization for the amplitude damping 
differs from the 
previous one in that the decay probability is now the 
same for each single qubit process, independently of 
the state of the system. Then, the decay probability 
for a states of the computational basis is 
proportional to the number of qubits in the up state.
This can be illustrated in the two-qubit 
case. In this example, the noise channel is described 
by the following unitary evolution formulae:
\begin{eqnarray}
|00\rangle_s |00 \rangle_e & \rightarrow & 
|00\rangle_s |00 \rangle_e, \nonumber \\
|01\rangle_s |00 \rangle_e & \rightarrow & 
\sqrt{1-p} \: |01\rangle_s |00 \rangle_e + 
\sqrt{p} \: |00\rangle_s |01\rangle_e, \nonumber \\
|10\rangle_s |00 \rangle_e & \rightarrow & 
\sqrt{1-p} \: |10\rangle_s |00 \rangle_e +
\sqrt{p} \: |00\rangle_s |10\rangle_e, \nonumber \\
|11\rangle_s |00 \rangle_e & \rightarrow & 
\sqrt{1-2p} \: |11\rangle_s |00 \rangle_e +
\sqrt{p} \nonumber \\
 & & (|10\rangle_s |01\rangle_e + |01\rangle_s |10\rangle_e).
\end{eqnarray} 
The Kraus operators $M_{\mu}^{(n)}$ for this model are given 
by the $n$-factor tensor product
\begin{equation}
M_{\mu}^{(n)}=\openone \otimes \ldots \otimes M_1 
\otimes \ldots \otimes \openone,
\label{KrausOps2}
\end{equation}
where $M_1$ is given by Eq.~(\ref{Kraus1})
and $\mu$ ($\mu=1,\ldots,n$) coincides with the 
position of $M_1$ in (\ref{KrausOps2}), that is $\mu$
singles out the qubit which decays from $|1\rangle$ to
$|0\rangle$. 
In the computational basis, the matrix representation
of the operator $M_0^{(n)}$ is given by 
\begin{equation}
[M_{0}^{(n)}]_{i,j} = \left\{
\begin{array}{ll}
1 & \; {\rm for}\; i=j=0, \\
 \sqrt{1-\sum_{l=0}^{n-1} i_l \frac{\Gamma dt}{\hbar}} & 
\;{\rm for}\; i=j \ne 0, \\ 
 & \\
 0  & \;{\rm otherwise}. 
\end{array} \right.
\end{equation}
The evolution of the pure state 
$\rho(t_0)=|1011\rangle\langle 1011|$ is different 
from what obtained in the previous many-qubit
damping model (see Eq.~(\ref{ampex}). We have
$$
\rho(t_0+dt)=\left(1-\frac{3\Gamma dt}{\hbar}\right)
|1011\rangle\langle 1011|+
\frac{\Gamma dt}{\hbar}(|0011\rangle\langle 0011|
$$
\begin{equation}
+|1001\rangle\langle 1001|+|1010\rangle\langle 1010|).
\end{equation}

\begin{figure}
\includegraphics[width=8.5cm,angle=0]{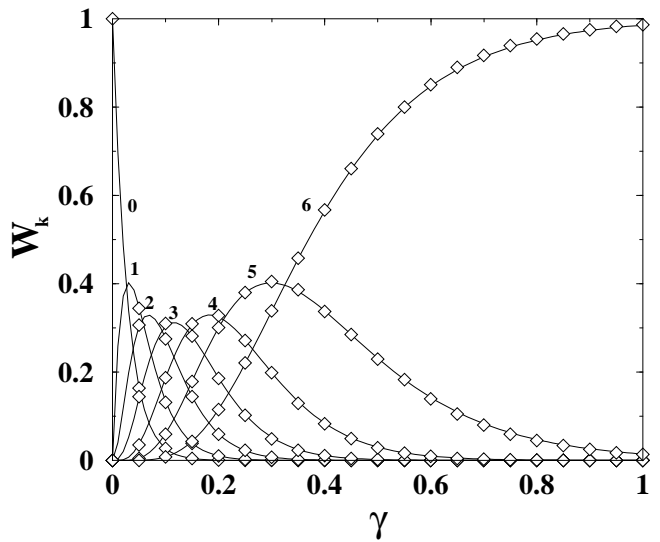}
\caption[]{\footnotesize Same as in Fig.~\ref{fig:cascade}, 
but for the noise model (\ref{KrausOps2}) and 
after evolution up to time $n dt$.} 
\label{fig:cascadech1}
\end{figure}

The cascade in the population of the different classes
$W_k$ is ruled by the following set of differential 
equations:
\begin{eqnarray}
\frac{dW_0}{dt} & = & -\frac{\Gamma}{\hbar} \; n_0 \; W_0, \nonumber \\
 & \vdots & \nonumber \\
\frac{dW_k}{dt} & = & \frac{\Gamma}{\hbar} \; (n_{k-1} \; W_{k-1} -  
n_k \; W_k), 
\nonumber \\
 & \vdots & \nonumber \\
\frac{dW_m}{dt} & = & \frac{\Gamma}{\hbar} \; n_{m-1} \; W_{m-1}
=\frac{\Gamma}{\hbar}{W_{m-1}},
\end{eqnarray}  
where $n_k$ is the number of qubits in the ``up'' state 
for the class $k$. Taking into account that $n_k=n_0-k=m-k$, the 
solutions are 
\begin{eqnarray}
W_k & = & \frac{n!}{n_k!} \; \sum_{i=0}^{k} 
\frac{(-1)^{(k-i)}}{i! \:(k-i)!} 
\; \exp{\left(-\frac{n_i \: \Gamma t}{\hbar}\right)},
\nonumber \\
W_m & = & 1 - \sum_{k=0}^{m-1} W_k.
\label{ADch1}
\end{eqnarray} 
In Fig.~\ref{fig:cascadech1}, we take the initial state 
$|2^n-1\rangle$ evolving it for a time $n \; dt$, for different 
values of the dimensionless damping rate $\gamma$. Again, only 
dissipation is considered and we have found a very good  
agreement with our analytical predictions. 

With this same idea we generalize the phase flip channel. 
The Kraus operators are given 
by the $n$-factor tensor product of \equ~(\ref{KrausOps2}), 
where now $\mu$ ($\mu=1,\ldots,n$) coincides with the 
position of the matrix $M_1$ of \equ~(\ref{Kraus2}), instead 
of \equ~(\ref{Kraus1}). 
The matrix representation of the Kraus operator $M_0^{(n)}$ in
the computational basis is
\begin{equation}
[M_{0}^{(n)}]_{i,j} = \left\{
\begin{array}{ll}
 \sqrt{1- n \frac{\Gamma dt}{\hbar}} & 
\;{\rm for}\; i=j, \\ 
 & \\
 0  & \;{\rm otherwise}. 
\end{array} \right.
\end{equation}
We study the stability of a given state vector subjected
to this decoherence channel. Namely, we compute the fidelity
$F={\rm Tr}[\rho_0 \rho(t)]$ of a $n$-qubit system,
for a random initial state $|\psi_0\rangle$, as a function 
of $\gamma=\Gamma dt/\hbar$ ($\rho_0=|\psi_0\rangle\langle\psi_0|$ 
is the density matrix for the initial state and 
$\rho(t)$ is the density matrix of the system at time $t$).
All components of the initial random state 
vector $|\psi_0\rangle$ are taken to be random complex numbers 
of modulus $1/\sqrt{2^n}$. 
A combinatorial calculation provides an exact closed expression for 
the fidelity in this case. We obtain  
\begin{equation}
F(t)=\frac{1}{2^n}+ \frac{n!}{2^n} \: \sum_{i=1}^n  
\frac{1}{i! \: (n-i)!} 
\: \exp{\left(-\frac{2 i \: \Gamma t}{\hbar}\right)}.
\label{eq:noisefidelity}
\end{equation}
Note that this formula does not give a simple exponential 
fidelity decay, but a superposition of exponential decays
with different rates, since the decay rate depends on the 
number of up spins in the different states of the 
computational basis.

\begin{figure}
\includegraphics[width=8.5cm,angle=0]{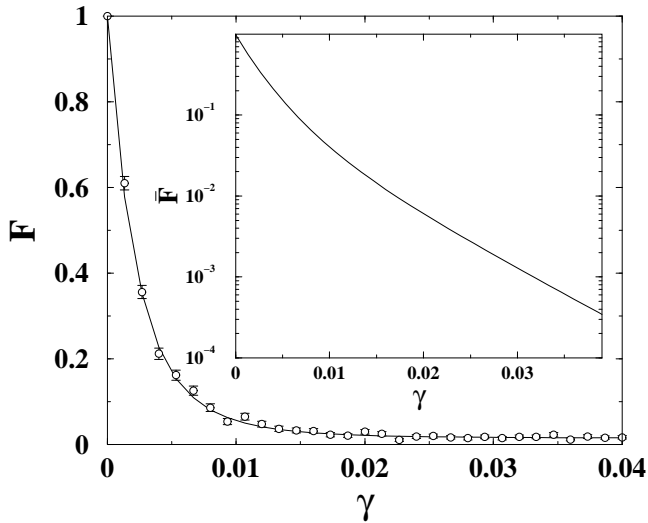}
\caption[]{\footnotesize Fidelity of a random initial 
state in terms of the dimensionless decay rate 
$\gamma=\Gamma dt/\hbar$, 
in a system with $n=6$ qubits, subjected to the generalized phase
flip channel up to time $2 n^2 dt$. 
The very good agreement between quantum 
trajectories simulations (circles with error bars) and 
theoretical predictions (solid line) is clearly seen. 
Here and in the following figures the error bars give the
size of the statistical error.
The inset shows the semi log version of the theoretical curve 
plot for the modified fidelity $\bar F=F-F_{\infty}$.}
\label{fig:fidelitych4}
\end{figure}

The agreement of the quantum trajectories simulations with this
theoretical formula is shown in Fig.~\ref{fig:fidelitych4}.
The semi log version of the theoretical curve is shown in the inset,
where we plot $F-F_\infty$, $F_\infty=1/2^n$ being the asymptotic
value of fidelity at $t=\infty$ or at $\Gamma=\infty$. 
We note that the asymptotic decay of $\bar F=F-F_{\infty}$
takes place with the lowest decay rate $\Gamma_1=2\Gamma/\hbar$.
of \equ~(\ref{eq:noisefidelity}).

We note that analytical formulae for fidelity decay can also be 
found for other special initial conditions.
For instance, \equ~(\ref{eq:noisefidelity}) remains valid
when the initial state is a random superposition of computational
basis states with up to $n_1$ spins up, provided that we replace
$n$ with $n_1$ in this equation.

In the following, we will apply the amplitude damping models 
to study the fidelity of a generalized teleportation protocol 
and the phase flip channel for the case of a quantum 
computer implementation of the baker's map.

									  
\section{Quantum teleportation}
\label{sec:tele}

\begin{figure}
\includegraphics[width=8.5cm,angle=0]{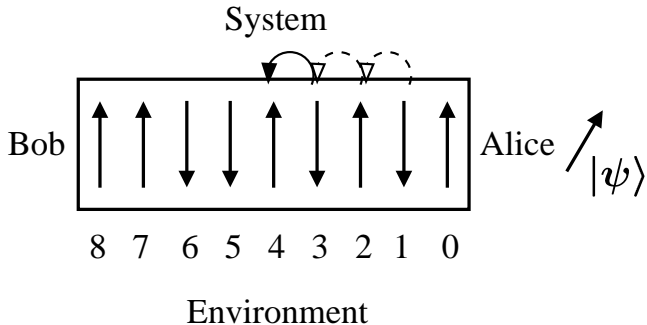}
\caption[]{\footnotesize Schematic drawing of the teleportation 
procedure that we study in the text. Alice sends one of the 
qubits of her pure Bell state to Bob. Meanwhile there is 
dissipation induced by the environment. 
In the figure, there is a chain of $n=9$ qubits and 
the third of the $n-2$ swap gates required by this quantum
protocol has been applied. Qubit $|\psi\rangle$ 
has to be teleported.}
\label{fig:scheme}
\end{figure}

Recently, there have been several publications 
focused on the investigation of fidelity of teleportation in the presence 
of a noisy environment \cite{Badziag,Bandyo,Oh,Verstraete}.
We study this problem in the situation presented in 
\cite{Car}, where a model of quantum teleportation through a 
noisy chain of qubits has been used. 
A schematic drawing of this quantum protocol is shown 
in Fig.~\ref{fig:scheme}.

Let us first recall this protocol in the ideal case, 
without environmental effects. 
We consider a chain of $n$ qubits, and assume that Alice 
can access the qubits located at one end of the chain, 
Bob those at the other end. Initially 
Alice owns an EPR pair (for instance we take the Bell 
state $|\phi^+\rangle = 
(|00\rangle+|11\rangle)/\sqrt{2}$), while the remaining 
$n-2$ qubits are in a pure state. Thus, the 
global initial state of the chain is given by  
\begin{equation}
\sum_{i_{n-1},\ldots,i_2} 
c_{i_{n-1},\ldots,i_2} 
|i_{n-1}\ldots i_2\rangle
\otimes \frac{1}{\sqrt{2}} (|00\rangle + |11\rangle),  
\label{phiin}
\end{equation}
where $i_k=0,1$ denotes the up or down state of the qubit $k$.
In order to deliver one of the qubits of the EPR pair to Bob,
we implement a protocol consisting of $n-2$ swap gates 
that exchange the states of pairs of qubits:  
$$
\sum_{i_{n-1},\ldots, i_2}
\frac{c_{i_{n-1},\ldots,i_2}}{\sqrt{2}} 
(|i_{n-1} \ldots i_2 0 0 \rangle + 
|i_{n-1} \ldots i_2 1 1 \rangle) 
$$
$$
\rightarrow
\sum_{i_{n-1},\ldots, i_2}
\frac{c_{i_{n-1},\ldots,i_2}}{\sqrt{2}} 
(|i_{n-1} \ldots 0 i_2 0 \rangle + 
|i_{n-1} \ldots 1 i_2 1 \rangle) 
\rightarrow 
$$
\begin{equation}
...\rightarrow 
\sum_{i_{n-1},\ldots, i_2}
\frac{c_{i_{n-1},\ldots,i_2}}{\sqrt{2}} 
(|0 i_{n-1} \ldots i_2 0 \rangle + 
|1 i_{n-1} \ldots i_2 1 \rangle). 
\end{equation}
After that, Alice and Bob share an EPR pair, and therefore 
an unknown state of a qubit ($|\psi\rangle=a|0\rangle +
b|1\rangle$) can be transferred from Alice to Bob by 
means of the standard teleportation protocol 
\cite{BennettBrassard}. 
In this work, we take random coefficients 
$c_{i_{n-1},\ldots,i_2}$, that is they
have amplitudes of the order of $1/\sqrt{2^{n-2}}$ (to assure 
wave function normalization) and random phases. This ergodic 
hypothesis models the transmission of a qubit through a chaotic 
quantum channel.  

We assume that our quantum protocol is implemented
by a sequence of instantaneous and perfect swap gates, 
separated by a time interval $\tau$, during which the 
corresponding noise channels introduce errors. 
This means that, using quantum trajectories, 
dissipation is implemented by means of  
``infinitesimal'' Kraus operators (see Sec.~\ref{sec:noise}), 
following the quantum jump numerical procedure outlined 
in Sec.~\ref{sec:QT:meqt}. 
We also assume that the only effect of the system's 
Hamiltonian $H_s$ is to generate the swap gates.

Let us call $\rho_k^{(n)}$ the density matrix of the whole chain
of $n$ qubits after $k$ swap gates.
Since the evolution of the density matrix in a time step
$dt$ is given, in the Kraus representation, by   
\equ~(\ref{eq:infsolution}), we can write the evolution
from $\rho_{k-1}^{(n)}$ to $\rho_k^{(n)}$ as follows: 
\begin{eqnarray}
\rho_k^{(n)}=U_{\rm sw}^{k,k+1} \: 
\left[ \sum_{\mu_1,...,\mu_k=0}^{\cal M} 
M_{\mu_k}^{(n)}(dt)\cdots M_{\mu_1}^{(n)}(dt) 
\right. 
\nonumber
\\ 
\rho_{k-1}^{(n)} \: 
\left.
(M_{\mu_1}^{(n)})^\dagger (dt) \cdots 
(M_{\mu_k}^{(n)})^\dagger (dt) 
\right]
 \: {U_{\rm sw}^{k,k+1}}^{\dag},
\end{eqnarray}
where $U_{\rm sw}^{ij}$ is the swap operator that exchanges 
the states of the qubits $i$ and $j$ and
$k=\tau/dt$ is the number of quantum noise
operations between two consecutive swap gates.
After $n-2$ swap gates, we obtain the final state 
$\rho_f^{(n)}=\rho_{n-2}^{(n)}$.

After computing the evolution of the initial 
state of the chain of $n$ qubits up to time $\Delta t= (n-2) \tau$, 
the standard teleportation protocol is implemented 
\cite{BennettBrassard}. The fidelity of teleportation 
is defined by 
\begin{equation}
F=\langle\psi|\rho_B|\psi\rangle,
\end{equation}
where $|\psi\rangle=a|0\rangle +b|1\rangle$ is the 
state to be teleported, and $\rho_B$ is the density matrix
of Bob's qubit at the end of the teleportation protocol, 
obtained after tracing over all the other qubits of the chain:
$\rho_B={\rm Tr}_{0,...,n-2} [\rho_f^{(n)}]$. 

In the quantum trajectories method, we compute the fidelity as 
\begin{equation}
F = \lim_{{\cal N}\to \infty}
\frac{1}{\cal N} 
\sum_{i=1}^{\cal N} \langle \psi | (\rho_B)_i 
| \psi \rangle,
\end{equation}
where $(\rho_B)_i$ is the reduced density matrix of 
Bob's qubit, obtained from the wave vector of the trajectory 
$i$ at the end of the quantum protocol. 
If the final state of the chain is 
$|\phi \rangle = \sum_{j=0}^{2^{n}-1} \alpha_j |j\rangle$ 
(an arbitrary state), the state of the 
whole system is 
\begin{eqnarray}
|\phi \rangle |\psi\rangle = & \frac{1}{\sqrt{2}} 
\sum_{j'=0}^{2^{n-1}-1} 
[(a \alpha_{2j'}+b \alpha_{2j'+1}) |j'\rangle |\phi^+ \rangle 
\nonumber \\
 & + (a \alpha_{2j'}-b \alpha_{2j'+1}) |j'\rangle |\phi^- \rangle .
\nonumber \\
 & + (b \alpha_{2j'}+a \alpha_{2j'+1}) |j'\rangle |\psi^+ \rangle  
\nonumber  \\
 & + (b \alpha_{2j'}-a \alpha_{2j'+1}) |j'\rangle |\psi^- \rangle],  
\end{eqnarray}
In the previous expression we have used the four (maximally 
entangled) Bell states 
($|\phi^\pm\rangle = (|00\rangle \pm |11\rangle)/\sqrt{2}$ and
$|\psi^\pm\rangle = (|01\rangle \pm |10\rangle)\sqrt{2}$), 
corresponding to the two least significant bits 
(the first bit of the chain and $|\psi\rangle$). 
Then, as required by the teleportation protocol, 
we perform a Bell measurement whose result determines one
out of possible unitary transformations acting on Bob's qubit.
Owing to quantum noise, Bob's qubit is entangled with 
the rest of the chain. Therefore, we must trace over these
qubits to obtain the reduced density matrix $\rho_B$ 
describing the state of Bob's qubit. 
If we measure $|\phi^+\rangle$ for instance, and define 
$\tilde \alpha_{j'}=1/\sqrt{2} \: (a \; \alpha_{2j'}+b \;
\alpha_{2j'+1})$, we arrive at the following expression:
\begin{equation}
\rho_{\rm B}= C
\left(
\begin{array}{ll}
\sum_{j'=0}^{D-1} | \tilde \alpha_{j'} |^2 & 
\sum_{j'=0}^{D-1} \tilde \alpha_{j'} 
\tilde \alpha_{j'+D}^* \\
\sum_{j'=0}^{D-1} \tilde \alpha_{j'}^* 
\tilde \alpha_{j'+D} &
\sum_{j'=D}^{2D-1} | \tilde \alpha_{j'} |^2 
\end{array}
\right),
\end{equation}
where $D=2^{n-2}$ and $C=1/\sum_{j'=0}^{2D-1} |\tilde \alpha_{j'}|^2$
is a normalization constant.
Then, the fidelity becomes 
\begin{eqnarray}
F= & C \big( |a|^2 \sum_{j'=0}^{D-1}|\tilde \alpha_{j'}|^2 + 
|b|^2 \sum_{j'=D}^{2D-1} |\tilde \alpha_{j'}|^2 +
\nonumber \\ 
 & + \sum_{j'=0}^{D-1} 2 \real{\tilde \alpha_{j'} \tilde 
\alpha_{j'+D}^* a^* b} \big).
\end{eqnarray}
In the special case $a=b=1/\sqrt{2}$, it reduces to
\begin{equation}
F= \frac{1}{2}+ C \sum_{j'=0}^{D-1} 
\real{\tilde \alpha_{j'} \tilde 
\alpha_{j'+D}^*} .
\label{fidteleportation}
\end{equation}
Similar expressions are obtained when the Bell measurement
gives outcomes $|\phi^{-}\rangle$, $|\psi^{+}\rangle$, or 
$|\psi^{-}\rangle$.

Let us now briefly discuss the case in which we 
directly evolve the density matrix 
of the whole system by means of the master 
equation (\ref{eq:lin2}). After the sequence of 
swap gates and quantum noise operations, we 
obtain the final density matrix $\rho_f^{(n)}$
for the $n$-qubit chain. 
The Bell measurement is performed by the operator
\begin{equation}
\frac{M_{\phi^+} (\rho_f^{(n)} \otimes \rho_{\psi}) 
M_{\phi^+}^{\dag}}
{{\rm Tr}[M_{\phi^+} (\rho_f^{(n)} \otimes \rho_{\psi})
M_{\phi^+}^{\dag}]},
\end{equation} 
where 
$\rho_\psi=|\psi\rangle\langle\psi|$ is the density matrix
describing the state of the qubit to be teleported and 
$M_{\phi^+}=\openone^{(n-1)} \otimes 
|\phi^+\rangle \langle\phi^+|$ 
($\openone^{(n-1)}$ is the identity for the remaining 
$n-1$ qubits of the chain).
Then, we take the partial trace and obtain the fidelity
$F={\rm Tr}(\rho_{\rm B} \rho_{\psi})=
\langle\psi |\rho_{\rm B} |\psi\rangle$. 
It is straightforward to see that the above procedure is 
equivalent to take the partial trace (over the mediating 
qubits of the chain) first and then proceed with low 
dimensionality calculations.

\begin{figure}
\includegraphics[width=8.5cm,angle=0]{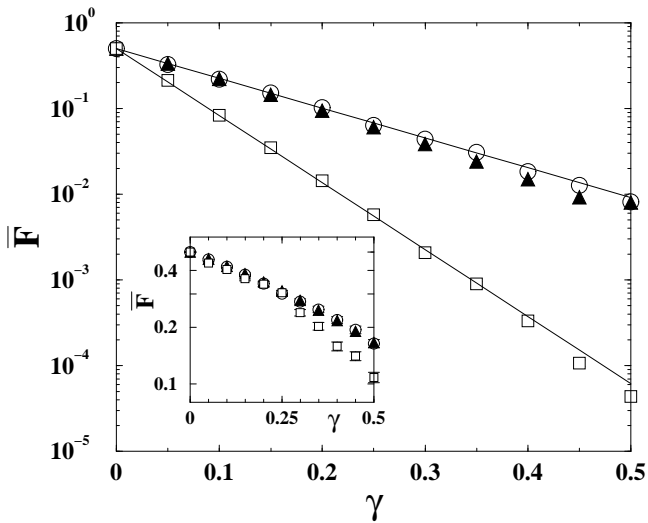}
\caption[]{Fidelity $\bar F=F-F_{\infty}$ ($F_{\infty}=1/2$) of 
the teleportation of 
the state $|\psi\rangle=(|0\rangle+|1\rangle)/\sqrt{2}$ as a 
function of the dimensionless damping rate 
$\gamma=\Gamma \tau/\hbar$, for the amplitude damping 
model (\ref{KrausOps2}). Circles and squares are the results of 
the quantum trajectories calculations for chains with $n=10$ and 
$n=20$ qubits, respectively. Triangles give the results 
of the density matrix calculations at $n=10$. 
Straight lines correspond to the theoretical result
of Eq.~(\ref{noise2q}). 
Inset: the same but for the noise model (\ref{amp1}).}
\label{telech01}
\end{figure}

We have investigated the effect of the two amplitude damping 
models described in Sec.~\ref{sec:noise} on the fidelity
of quantum teleportation.
In Fig.~\ref{telech01}, we show the results of our numerical
simulations for the special case in which the state to be
teleported is $|\psi\rangle=(|0\rangle+|1\rangle)/\sqrt{2}$.
In this case, in the limit of infinite chain ($n\to\infty$) 
or of large damping rate, 
the density matrix $\rho_{\rm B}$ describing the state
of Bob's qubit becomes $\rho_{\rm B}=|0\rangle\langle 0|$. 
Thus, the asymptotic value of fidelity is given by 
$F_\infty=1/2$ and we plot the values of 
$\bar F=F-F_{\infty}$, for the noise models 
(\ref{amp1}) and (\ref{KrausOps2}), corresponding to the inset
and the main figure in Fig.~\ref{telech01}, respectively.
For both cases we have checked the accuracy of the quantum trajectory 
simulations by reproducing the results with direct density matrix 
calculations. This was possible only up to $n=10$ qubits.
The exponential decay of fidelity in the case in which quantum
noise is modeled by Eq.~(\ref{KrausOps2}) is in agreement with the
theoretical formula 
\begin{equation}
F=\frac{1}{2}+\frac{1}{2}\exp{\left(-\frac{\Gamma t}{\hbar}\right)}=
\frac{1}{2}+\frac{1}{2}\exp(-\gamma k),
\label{noise2q}
\end{equation}
where $\gamma=\Gamma \tau /\hbar$ is the dimensionless 
damping rate and $k=t/\tau=n-2$ measures the time in 
units of quantum (swap) gates.
To derive this theoretical formula, we observe that this 
quantum noise model does not generate entanglement 
between the two qubits of the Bell pair 
and the others qubit of the chain. Therefore, 
it is sufficient to study the evolution of the Bell state 
$|\phi^+\rangle\langle \phi^+|$ under the noise model 
(\ref{KrausOps2}). This evolution takes place inside
a two-qubit Hilbert space of dimension $4$ and its 
exact solution is given by Eq.~(\ref{noise2q}).  
On the contrary, the quantum noise model (\ref{amp1})
entangles these two qubits with the rest of 
the chain. In this case, the fidelity decay is not 
exponential. Unfortunately, we could not provide an 
analytical derivation of $F(t)$ in this case.
 
\section{The quantum baker's map}
\label{sec:Baker}

The quantum algorithm for the simulation of the quantum baker's
map has been proposed by Schack \cite{Sch} and recently 
implemented by means of a three-qubit NMR-based quantum processor
\cite{cory}, where the fidelity of the quantum computation of
the baker's map in one map step was measured. 

The baker's transformation is one of the prototype models of classical and
quantum chaos \cite{Bala}. It maps the unit square 
$0\le{q,p}<1$ onto itself according to
\begin{eqnarray}
q_{k+1} & = & 2 q_k - [2 q_k],\nonumber \\
p_{k+1} & = & (p_k+[2q_k])/2, 
\label{bakermap}
\end{eqnarray}
where $[x]$ stands for the integer part of $x$ and the index $k$ 
denotes the number of map iterations. The action of this map 
corresponds to compressing the unit square in the $p$ direction and
stretching it in the $q$ direction, then cutting it along the $p$
direction, and finally stacking one piece on top of the other
(similarly to the way a baker kneads dough). Note that map 
(\ref{bakermap}) is area preserving.
The baker's map is a paradigmatic model of classical chaos.
Indeed, it exhibits sensitive dependence on initial conditions, 
which is the distinctive feature of classical chaos: any small 
error in determining the initial conditions is 
amplified exponentially in time. In other
words, two nearby trajectories separate exponentially, with a rate 
given by the maximum Lyapunov exponent $\lambda=\ln 2$.

The baker's map can be quantized following \cite{Bala}. 
We introduce the position ($q$) and momentum ($p$) operators, 
and denote the eigenstates of these operators by $|q_j\rangle$ and 
$|p_k\rangle$, respectively. The corresponding eigenvalues
are given by $q_j=j/N$ and $p_k=k/N$, with $j,k=0,\dots,N-1$, $N$ 
being the dimension of the Hilbert space. Note that, to fit $N$ 
levels onto the unit square, we must set $2\pi \hbar= 1/N$. 
Therefore, the effective Planck's constant of the 
system is $\hbar_{\rm eff}\propto 1/N$.
We take $N=2^n$, where $n$ is the number of qubits used to simulate 
the quantum baker's map on a quantum computer. 
Note that $\hbar_{\rm eff}$ drops exponentially with the
number of qubits and therefore the semiclassical regime
$\hbar_{\rm eff}\ll 1$ can be reached with a small number of qubits.
The transformation between the position basis
$\{|q_0\rangle,\dots,|q_{N-1}\rangle\}$ and the momentum basis
$\{|p_0\rangle,\dots,|p_{N-1}\rangle\}$ is performed by means of
a discrete Fourier transform $F_n$, defined by the matrix elements
\begin{equation}
  \langle q_k | F_n | q_j \rangle \equiv
  \frac1{\sqrt{2^n}} \, \exp\!\left( \frac{2\pi ikj}{2^n} \right) .
  \label{dftbaker}
\end{equation}
It can be shown \cite{Bala} that the quantized baker's map may 
be defined by the transformation 
\begin{equation}
  |\psi_{k+1}\rangle =
  B \, |\psi_k\rangle = 
  F_n^{-1}
  \left[
    \begin{array}{cc}
      F_{n-1} & 0 \\
      0 & F_{n-1}
    \end{array}
  \right] 
  |\psi_k\rangle,
\label{baker1k}
\end{equation}
where $|\psi_k\rangle$ denotes the wave vector of the system
after $k$ map steps,  the matrix elements are to be  
understood relative to the position basis
$\{|q_j\rangle\}$ and $F_{n-1}$ is the discrete Fourier transform,
defined by Eq.~(\ref{dftbaker}).

Since the discrete Fourier transform can be calculated on 
a quantum computer using $O(n^2)$ elementary gates
(see, e.g., Ref.~\cite{Chuangbook}), the simulation of one 
step of the baker's map requires $O(n^2=(\log N)^2)$ gates 
\cite{Sch} (more precisely, we need $(n-1)^2$ controlled phase-shift
gates and $2n-1$ Hadamard gates). Therefore, it is exponentially 
faster than the best known classical computation, which is based on
the fast Fourier transform and requires $O(N\log N)$
gates.   

In this section, we investigate the fidelity of the quantum computation
of the baker's map in the presence of quantum noise.
We consider the phase flip noise channel, generalized to the 
$n$-qubit case as discussed in Sec.~\ref{sec:noise}. We  
take an initial state $|\psi_0\rangle$ with amplitudes of the 
order of $1/\sqrt{2^n}$ and random phases. 
We perform the forward evolution of the baker's map 
up to time $k$ (this evolution is driven, in the noiseless case, 
by the unitary operator $B^k$, with $B$ given in Eq.~\ref{baker1k}), 
followed by the $k$-step backward evolution (represented by 
the operator $(B^\dagger)^k$). Due to quantum noise, the initial state 
$|\psi_0\rangle$ is not exactly recovered and the final state
of the system is described by a density matrix $\rho_f$.
The fidelity of quantum computation is given by 
$F=\langle \psi_0 |\rho_f |\psi_0 \rangle$.

We are able to work out the following theoretical formula for 
the decay of fidelity induced by phase flip noise:
\begin{equation}
F = \exp\left(-n\gamma N_g \right)=
\exp(-2\gamma n^3 k),
\label{eq:Bakerfidelity}
\end{equation}
where $\gamma=\Gamma \tau/\hbar$ is the dimensionless decay rate
(again, $\tau$ denotes the time interval between elementary 
quantum gates) and 
$N_g=2 n^2 k$ is the total number of elementary quantum gates 
required to implement the $k$ steps forward evolution of the 
baker's map, followed by $k$ step backward.
To derive this formula, we first of all note that 
\equ~(\ref{eq:noisefidelity}), obtained in the absence of 
any quantum gate operation, gives, at short times, 
the exponential decay
\begin{equation}
F(t)= \exp(-\frac{<\Gamma>t}{\hbar}).
\end{equation} 
Here the decay rate $<\Gamma>$ is obtained after averaging
all the decay rates that appear in Eq.~(\ref{eq:noisefidelity}):
\begin{equation}
<\Gamma> = \frac{n!}{2^n} \: \sum_{i=0}^n  
\frac{1}{i! \: (n-i)!} 
2 i \Gamma = n \Gamma.
\end{equation}
The effect of the chaotic dynamics of the baker's map is to 
induce a fast decay of correlations, so that any memory of the 
initial state is rapidly lost and the fidelity decay 
remains exponential also at long times.
Therefore, the condition for the validity of 
formula (\ref{eq:Bakerfidelity}) is that the randomization 
process introduced by chaotic dynamics takes place in a time
scale shorter than the time scale for fidelity decay.

We can determine from Eq.~(\ref{eq:Bakerfidelity}) the time 
scale up to which a reliable quantum computation of the 
baker's map evolution in the presence of the phase flip
noise channel is possible even without quantum error
correction. The time scale $k_f$ at which $F$ drops below 
some constant $A$ (for instance, $A=0.9$) is given by 
\begin{equation}
k_f=-\frac{\ln A}{2\gamma n^3}.
\end{equation}
The total number of gates that can be implemented up to this 
time scale is given by 
\begin{equation}
(N_g)_f= 2 n^2 k_f =-\frac{\ln A}{\gamma n}.
\label{ngrel}
\end{equation}
 
\begin{figure}
\includegraphics[width=8.5cm,angle=0]{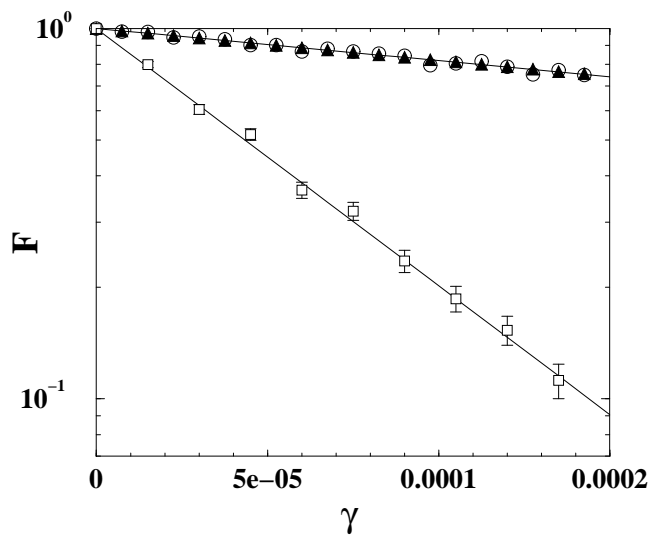}
\caption[]{Semi log plot of the fidelity as a function 
of the dimensionless decay rate $\gamma$, 
for the baker's map after one map step in the 
presence of the phase flip channel described in the text. 
Circles and squares correspond to quantum trajectories 
simulations with ${\cal N}=500$ trajectories, at $n=10$ 
and $n=20$ qubits, respectively. Triangles give the 
results obtained by direct computation of the 
density matrix evolution at $n=10$. 
Solid lines stand for the theoretical prediction 
of \equ~(\ref{eq:Bakerfidelity}), namely
$F(\gamma)=\exp(-2\gamma n^3)$.}
\label{B:fid}
\end{figure}

\begin{figure}
\includegraphics[width=8.5cm,angle=0]{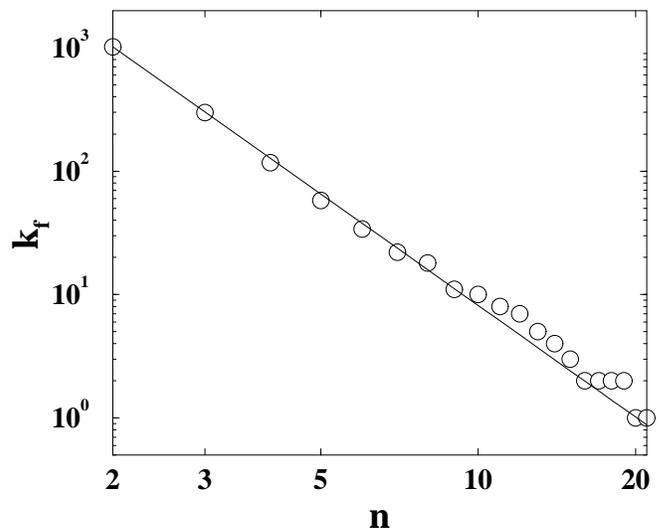}
\caption[]{Logarithmic plot of the time scale $k_f$
(measured in number of steps of the baker's map) 
required to reach a fidelity value of $F=0.9$, 
as a function of the number $n$ of qubits, 
for $\gamma\approx 6.47\times 10^{-5}$
(this value has been chosen by demanding that $F=0.9$
after one forward/backward step of the baker's map
at $n=21$).
The straight line gives the theoretical curve 
$T(n)=C/n^3$, with $C=-\ln(0.9)/(2\gamma)\approx 8.14$.} 
\label{B:fid2}
\end{figure}

Our theoretical expectations are confirmed by the
numerical data of Figs.~\ref{B:fid} and \ref{B:fid2}. 
In Fig.~\ref{B:fid}, we show the fidelity after 
one map step, as a function of the dimensionless damping rate 
$\gamma=\Gamma \tau/\hbar$.
The numerical data of this figure show that the fidelity 
drops exponentially with $\gamma$, in excellent agreement with 
Eq.~(\ref{eq:Bakerfidelity}). 
We point out that our theory predicts not only the 
exponential fidelity decay but also the numerical value of the 
decay rate.
Finally, we show in Fig.~\ref{B:fid2} the number $k_f$ of 
forward/backward map steps required to reach a fixed 
value $A$ of the fidelity ($F=A=0.9$), as a function of 
the number $n$ of qubits. This time scale decays as a power 
law, $k_f = C/n^3$, in agreement with Eq.~(\ref{eq:Bakerfidelity}).
Again, our theory predicts also the value of the prefactor 
$C$, in excellent agreement with numerical data.

As we have seen above, the number of gates that 
can be reliably implemented without quantum error correction
drops only polynomially with the number of qubits, 
$(N_g)_f\propto 1/n$ (see Eq.~\ref{ngrel}). 
We would like to stress that this dependence 
should remain valid also in other 
environment models that allow only one qubit at a time 
to perform a transition, like the other noise channels
descussed in Sec.~\ref{sec:noise}.
Furthermore, we note that the time scales for fidelity 
decay derived in this section and confirmed in the 
baker's map model, are expected to be valid for any quantum 
algorithm simulating dynamical systems in the regime 
of quantum chaos.



\section{Conclusion}
\label{sec:conclusion}

We have studied two quantum protocols, i.e. a teleportation 
scheme through a chain of qubits and a quantum algorithm for  
the quantum baker's map. We have modeled different sorts of 
environments in order to get a deeper insight of the stability 
of quantum computation when different interactions with the environment 
taken into account. Two kinds of generalized amplitude damping models 
have been considered for the teleportation scheme, giving very different 
behaviors. After a theoretical analysis we were able to understand the 
origin of these differences. This reveals the importance of the details 
of the environmental models in assessing the operability bounds of 
quantum processors. 
In the case of the baker's map simulation, we have chosen the 
phase flip type of noise and we could verify our theoretical predictions
for fidelity decay. 
The results of this paper show that quantum trajectories are a very
valuable tool for simulating noise processes in quantum information 
protocols with a high degree of efficiency.   

\begin{acknowledgments}

This work was supported in part by the EC contracts 
IST-FET EDIQIP and RTN QTRANS, the NSA and ARDA under
ARO contract No. DAAD19-02-1-0086, and the PRIN 2002 
``Fault tolerance, control and stability in
quantum information precessing''.

\end{acknowledgments}



\end{document}